\documentstyle[aps,twocolumn]{revtex}

\begin{document}

\newcommand{\be}{\begin{equation}}
\newcommand{\ee}{\end{equation}}
\newcommand{\bea}{\begin{eqnarray}}
\newcommand{\eea}{\end{eqnarray}}
                        
\draft
\tighten

\title{Atom cooling and trapping by disorder}

\author{Peter Horak, Jean-Yves Courtois, and Gilbert Grynberg}
\address{Laboratoire Kastler-Brossel, Ecole Normale Supérieure, 24 rue
Lhomond, 75231 Paris Cedex 05, France.}
\date{\today}
\maketitle

\begin{abstract}
We demonstrate the possibility of three-dimensional cooling of neutral atoms 
by illuminating them with {\em two\/} counterpropagating laser beams
of mutually orthogonal linear polarization, where one of the lasers is a
speckle field, i.e.\ a highly disordered but stationary coherent light field. 
This configuration gives rise to
atom cooling in the {\em transverse\/} plane via a Sisyphus cooling mechanism
similar to the one known in standard two-dimensional
optical lattices formed by several plane laser waves. 
However, striking differences occur in the spatial
diffusion coefficients as well as in local properties of the trapped atoms.
\end{abstract}

 
\narrowtext

\narrowtext


\section{Introduction}

The study of speckle laser patterns, as created when a highly coherent
light beam is transmitted through or reflected from an object with a
surface which is rough on the scale of the laser wavelength, was initiated
many years ago using electromagnetic theory \cite{elmag} and statistical
methods \cite{Goodman,statistical}. Since then, this subject has raised
more and more interest, resulting in a vast development of the theory (for
a survey of the most recent results, see e.g.\ \cite{newtheory}) as well
as experimental achievements. These give rise to important applications of
laser speckles in various fields of science, such as
the speckle reduction in imagery \cite{imagery}, roughness measurements in
material science \cite{roughness},
or applications in bio-physics \cite{biophysics}.

The scope of the work presented here is to investigate the application of
such speckle laser fields in the context of laser cooling of neutral atoms.
We are especially interested in the disordered analog of the so-called
optical lattices formed by
several laser plane waves, which yield periodic optical potentials and 
have been demonstrated experimentally to give
rise to efficient laser cooling by a Sisyphus-type mechanism 
\cite{Lattices1,Lattices2,Lattices3}. The basic principle of this cooling
scheme is that the atoms loose kinetic energy by running up potential hills
from where they are optically pumped into lower lying potential wells. 
Recent experiments have also demonstrated a similar cooling scheme in the
case of laser configurations forming quasi-periodic optical
lattices \cite{Quasicrystals}, which can be viewed as an intermediate
regime between the standard periodic lattices and the completely
disordered patterns obtained from speckle light fields.

We consider two counterpropagating laser fields of mutually orthogonal
polarization, where one of the beams is a speckle field. According to the
randomly distributed
phase and intensity gradients of the speckle field in all dimensions we find
three-dimensional Sisyphus cooling even with this one-dimensional beam
configuration. Moreover, interesting
transport phenomena are found in this case, e.g., a large difference in
the spatial diffusion of the atoms even in parameter regimes where the
steady-state temperatures are of the same order of
magnitude for the longitudinal and the transverse directions.
Other effects, such as
local radiation pressure forces, arise also from the fact that amplitude and
phase of the speckle field are essentially independent.

The paper is organized as follows. In Sec.~\ref{sec:qualitative} we
qualitatively discuss our proposed setup and the basic physical processes
which give rise to the cooling mechanism. Furthermore we discuss the
dependence of the steady-state temperature on various system parameters.
In Sec.~\ref{sec:theory} we present the details of the theory and outline
the numerical methods,
especially the semiclassical Monte-Carlo simulations, to obtain
the results. Sec.~\ref{sec:numerics} is attributed to the analysis of the 
numerically obtained results such as temperatures, spatial diffusion
coefficients and local effects in the light field. Finally, in
Sec.~\ref{sec:longitudinal} we discuss the differences of the cooling
in the laser propagation direction and in the transverse plane.


\section{Qualitative discussion}
\label{sec:qualitative}

Throughout this paper we will discuss the simple situation of a single atom
with a ground state of total angular momentum $F=1/2$
and an excited state with $F'=3/2$ interacting with two counterpropagating
laser fields of orthogonal polarization. In the case of both laser fields
being plane waves this gives rise to well-known sub-Doppler cooling
mechanisms in one dimension (1D) \cite{Polgrad}. 
Because of the modulation
of the optical potential, trapping of atoms in the longitudinal direction
has been predicted \cite{Pgrad12a} and observed \cite{Pgrad12b}. In
contrast, the atoms are free in the transverse plane.

In the situation discussed here, one of the laser fields is replaced by a
speckle field, i.e., by a highly disordered, but nevertheless stationary and
coherent light field. Such a speckle field can be easily generated
experimentally, for instance, by introducing a diffusor into the path of a
laser plane wave. The resulting light field shows highly disordered
intensity and phase distributions (for a discussion of the statistical
properties of speckle fields see e.g.\ \cite{Goodman}). An example of a
computer-generated speckle
field on a discrete spatial grid is shown in Fig.~\ref{fig:speckle}.

We have investigated two different possible setups for the polarizations
of the counterpropagating light fields, namely the case of two mutually
orthogonal linearly polarized fields (lin$\perp$lin configuration) and the
case of two
opposite circular polarizations ($\sigma_+ - \sigma_-$ configuration).

For the lin$\perp$lin configuration one finds in the laser propagation
direction essentially the usual 1D Sisyphus cooling mechanism \cite{Polgrad}
apart from the different spatial variation of intensity and phase of the
speckle field as compared to a plane wave. Additionally, the properties of
the speckle field give rise to Sisyphus cooling of the atom in the
transverse plane, since the phase and intensity modulations in these
directions lead to spatially varying light shifts and optical pumping
rates of the Zeeman sublevels of the atomic ground state. Thus, the two
counterpropagating laser beams (one plane wave and one speckle field)
allow for {\em three-dimensional cooling} of the atom, in sharp contrast
to the 1D lin$\perp$lin optical lattices used so far.

Sisyphus cooling in the plane orthogonal to the laser propagation
direction works analogously in the $\sigma_+ - \sigma_-$ configuration.
However, for an $F=1/2$ to $F'=3/2$ atomic transition, Sisyphus cooling
in the longitudinal direction is expected to be much less efficient
since the optical potentials of the ground
state sublevels change on a much larger length scale in this direction
(if the typical length size of the speckle in the transverse
direction is $d_{sp}$, the size along the propagation (longitudinal) axis is
$d_{sp}^2/\lambda$, where $\lambda$ is the laser wavelength).
Furthermore, the $\sigma_+ - \sigma_-$ cooling scheme \cite{Polgrad} is
efficient only if $F \ge 1$.

In these two configurations the most striking difference to the
configuration of counterpropagating plane waves
will be found in the transverse plane. Hence in the
remainder of this paper we will mainly restrict ourselves to a 2D model 
of cooling in the transverse directions and will only briefly discuss
the changes induced by a model including the longitudinal direction
in Sec.~\ref{sec:longitudinal}.

As an example we plot the steady-state temperature reached in this 2D
subsystem versus the optical potential depth $\hbar \Delta^\prime$ for
fixed optical pumping rate $\gamma'$ in Fig.~\ref{fig:temp}, where 
$\Delta'$ and $\gamma'$ for the speckle field are defined with respect
to the mean intensity of the beam.
The numerical method to obtain these results will be described in detail in 
Sec.~\ref{sec:theory}. As our first and most important result we note that,
as expected from the previous discussions, the atoms reach a steady-state
temperature in the transverse plane. Secondly, the general behavior of this
steady-state temperature as a function of the optical potential depth
resembles very much the case of the standard optical lattices consisting
of plane laser waves, i.e., for large values of $\Delta^\prime$ one finds
a linear dependence and hence
the steady-state kinetic energy becomes constant relative to the depth of
the optical potentials in this limit. On the other hand, the temperature
increases dramatically with small, decreasing values of $\Delta^\prime$
and, finally, below a certain threshold value of $\Delta^\prime$ no
steady-state temperature is achieved. Consequently, one finds the lowest
absolute values for the steady-state temperature for intermediate values
of the optical potential depth.

According to this similarity with standard optical lattices we may roughly
estimate the steady-state temperature of the atoms in the speckle field from 
the formula 
\be
   k_B T=\frac{\overline{D}}{\overline{\alpha}},
\label{eq:temp}
\ee
where $\overline{D}$ is a mean momentum diffusion coefficient and 
$\overline{\alpha}$ a mean friction coefficient \cite{DalibardCohen}. Note
that this formula only holds in the absence of atomic localization, which
is not the case here as we will see later, but nevertheless gives a useful
order of magnitude for the temperature.
The friction coefficient will be of the order of 
\be
   \overline{\alpha} \sim \hbar k^2\left(\frac{\lambda}{d_{sp}}\right)^2 
                  \frac{\Delta}{\Gamma},
\label{eq:fric}
\ee
where $\Delta$ is the detuning of the laser from the atomic resonance
frequency, $\Gamma$ is the natural linewidth of the excited state, and
$d_{sp}$ is the mean distance of neighboring intensity maxima along a 1D
cut through the speckle field, i.e.\ the typical length scale
of the speckle field. Eq.~(\ref{eq:fric}) is obtained from the well-known
expression for standard lattices \cite{Polgrad,LesHouches}
where we have only replaced the factors of $\lambda$ by $d_{sp}$, which
accounts for the different typical length scale.

The momentum diffusion coefficient can be guessed as
\be
  \overline{D} = \overline{D}_{dip} + \overline{D}_{se} \sim
     \hbar^2 k^2 \left( \frac{\lambda}{d_{sp}} \Delta' \right)^2 
        \frac{1}{\gamma'}
     + \gamma' \hbar^2 k^2,
\ee
which contains the diffusion $\overline{D}_{dip}$ due to the fluctuating
dipole force and the diffusion $\overline{D}_{se}$ due to the recoil of
the spontaneously emitted photons. $\overline{D}_{dip}$ is approximately
given by the square of a typical force, 
$\hbar k (\lambda/d_{sp}) (\Delta/\Gamma)$, times a typical time, 
$1/\gamma'$, and $\overline{D}_{se}$ has the same form as for a standard
lattice. Hence the estimated temperature (\ref{eq:temp}) reads
\be
  \frac{k_B T}{\hbar \Delta'} \sim 1 + 
        \left(\frac{d_{sp}}{\lambda}\right)^2 
        \left( \frac{\Gamma}{\Delta} \right)^2.
\label{eq:temp2}
\ee
This formula exactly predicts the qualitative behavior of the temperature
in Fig.~\ref{fig:temp}, i.e.\ the linear increase for large values of 
$\Delta^\prime$ and a rapid increase for very small values.
It also shows that the lowest value of $T$ ($=T_{min}$) is achieved for 
$\Delta'/\gamma'=d_{sp}/\lambda$ and that
$k_B T_{min}=2\hbar\gamma' d_{sp}/\lambda$. For a fixed optical pumping rate
$\gamma'$, the minimum temperature is thus
expected to increase linearly with the speckle size.

Finally, one notes from Fig.~\ref{fig:temp} that, although both the 
lin$\perp$lin and the $\sigma_+ - \sigma_-$ configuration give rise to
cooling, the lin$\perp$lin setup is more efficient and hence gives rise to
lower temperatures. For the chosen parameters the difference in temperature
between these two
setups is about a factor of two in the vicinity of the threshold value for
the optical potential depth and is smaller for larger values of $\Delta'$.

Another important difference between the laser cooling inside a speckle
field as discussed here and standard optical lattices created by several
plane waves lies in the tunability of the typical length scale. For our
setup the mean speckle grain size $d_{sp}$ can be changed continuously by
changing the position of the diffusor
which creates the speckle field out of a plane laser wave.

As an example, Fig.~\ref{fig:specklesize} shows the dependence of the
steady-state temperature on $d_{sp}$ for the lin$\perp$lin as well as for 
the $\sigma_+ - \sigma_-$ configuration. One clearly sees an increase of
the temperature for larger speckle sizes. 
Eqs.~(\ref{eq:fric})-(\ref{eq:temp2}) explain these feature since we note
that the friction force decreases as $1/d_{sp}^2$, but only the dipole
diffusion term $\overline{D}_{dip}$ has the same dependence, whereas the
sponaneous emission term $\overline{D}_{se}$ remains constant. Hence the
latter contribution of the total diffusion starts to
dominate for large speckle sizes (more precisely for 
$d_{sp}/\lambda \gg \Delta'/\gamma'$), 
and hence the temperature is expected to
increase quadratically with the speckle size.
However, for the parameters chosen in Fig.~\ref{fig:specklesize} 
($\Delta'/\gamma'=15$ ) the temperature increases only by a factor of two,
if the speckle grain size is increased from about $\lambda$ to $12\lambda$.
But as we will discuss later
in Sec.~\ref{sec:numerics}, the increase, for instance, of the spatial
diffusion coefficient, which is equal to 
$\overline{D}/\overline{\alpha}^2$ in a
certain range of parameters \cite{Hodapp}, is much larger.

Again the qualitative behavior of the $\sigma_+ - \sigma_-$ configuration
is the same as that of the lin$\perp$lin configuration. Thus in the
remainder of the paper we will only focus on the latter situation, since
it is more appropriate for a 3D cooling mechanism as discussed previously.


\section{Theoretical model}
\label{sec:theory}

In this section we give a more detailed discussion of the mathematical
model and the numerical methods used to obtain the results presented
in this paper.

\subsection{Speckle fields}

The first step for all the numerical treatments is to create a speckle
field. This can be easily implemented on a computer following the 
procedure described in the work of Huntley \cite{Huntley}, which we will
briefly outline in the following.

Let us denote the electric field at the diffusor by $E_1(x,y)$
and the electric field at the plane chosen for the numerical simulations by 
$E_2(x,y)$. The two fields are related according to the Huygens-Fresnel
principle in the Fresnel approximation by
\bea
E_2(x,y) & = & \frac{1}{\lambda l}\exp \left[ -i \frac{\pi}{\lambda l} 
          (x^2 + y^2) \right] \nonumber \\
   & \times &  \int dx_1 dy_1
	 E_1(x_1,y_1) 
	   \exp \left[ -i \frac{\pi}{\lambda l} (x_1^2 + y_1^2) \right] 
	   \nonumber \\
   & \times & \exp \left[ i \frac{2\pi}{\lambda l} (x x_1 + y y_1) \right],
\eea
where $l$ denotes the distance between the diffusor and the observation
plane. Hence, if we define
\bea
E_1^\prime (x,y) &=& E_1 (x,y) 
            \exp\left[ -i\frac{\pi}{\lambda l}(x^2 + y^2) \right],\\
E_2^\prime (x,y) &=& E_2 (x,y)
            \exp\left[ i\frac{\pi}{\lambda l}(x^2 + y^2) \right],
\eea
these two quantities are related by a Fourier transform.

Following \cite{Goodman} the real and imaginary part of 
$E_2^\prime (x,y)$ at any point $(x,y)$ are independent Gaussian random
variables. Thus the numerical construction of a speckle field starts by
filling the real and imaginary parts of $E_2^\prime (x,y)$ on a discrete
spatial grid of $N\times N$ points with random numbers from a Gaussian
distribution of zero mean and unit standard deviation. 
The grid size $N$ must be chosen in such a way that it contains a
reasonable large number of speckle grains, but nevertheless the
discretized field must be smooth enough in order not to introduce large
numerical errors. In practice we have used values of $N$ between 
64 and 256 for our calculations. 

The effect of the finite size of the diffusor is implemented by
transforming $E_2^\prime (x,y)$ into $E_1^\prime (x,y)$,
multiplying the latter quantity by a window function $W(x,y)$ which
assumes unity inside the diffusor and vanishes otherwise, and transforming
the result back into a final $E_{2, \mbox{\scriptsize fin}}^\prime (x,y)$,
i.e.,
\be
E_{2, \mbox{\scriptsize fin}}^\prime = 
   {\cal F}^{-1}(W {\cal F}(E_2^\prime)),
\ee
where ${\cal F}$ denotes the two-dimensional Fourier transform.

\subsection{Semiclassical model of atomic dynamics}
\label{sec:semiclass}

As in most of the previous theoretical works on the Sisyphus cooling
mechanism, including semiclassical \cite{Polgrad,LesHouches,Castin}
as well as quantum treatments \cite{QuantSisy,Levy},
we consider the simple case of an atom with a ground state of
angular momentum $F=1/2$ and an excited state of angular momentum $F'=3/2$.
Furthermore we restrict ourselves to a 2D model and to the case of low
atomic saturation, where we can adiabatically eliminate the excited state
of the atom. The time evolution of the atomic density operator restricted
to the ground state manifold is then governed by the master equation
\be
   \dot{\rho} = -\frac{i}{\hbar} [H,\rho] + L\rho,
   \label{eq:master}
\ee
with the Hamiltonian
\be
   H = \hat{p}^2/2m + \hbar \Delta' V(\hat{x})
\ee
and the decay and recycling term
\bea
   L\rho &=& \gamma^\prime/2 \Big[ 
      -V(\hat{x})\rho - \rho V(\hat{x}) \nonumber \\
      &+& 2 \sum_{\sigma}\int d^2 q N(q)
      e^{-iq\hat{x}} B_{\sigma}^{\dagger}(\hat{x}) \rho
      B_{\sigma}(\hat{x}) e^{iq\hat{x}},
   \Big]
\eea
where the sum goes over the polarization and the integral over the wave
vector of the spontaneously emitted photon projected into 2D.
The optical potential depth is $\hbar \Delta^\prime$, the optical pumping
rate is $\gamma^\prime$, and the atomic transition operators 
$B_{\sigma}(\hat{x})$
and the optical potential operator $V(\hat{x})$ are defined as
\bea
   B_{\sigma}(\hat{x}) & = & \left[ 
          \sum_{\mu} E_{\mu}(\hat{x})^\dagger A_{\mu}
       \right] A_{\sigma}^\dagger, \\
   A_{\sigma} & = & \sum_{m',m} \langle F',m' |1,\sigma; F,m\rangle
       |F',m'\rangle \langle F,m|, \label{eq:Aop} \\
   V(\hat{x}) & = & \sum_{\sigma} B_{\sigma}(\hat{x}) 
                    B_{\sigma}^\dagger(\hat{x}),
\eea
where $E_{\sigma}(x)$ gives the spatial dependence of the 
$\sigma$-polarized laser light and where we have made use of the
Clebsch-Gordan coefficients in Eq.~(\ref{eq:Aop}).

In order to derive a semiclassical theory suitable for Monte-Carlo
simulations we rewrite the master equation (\ref{eq:master}) in the
Wigner representation defined by
\be
   W(x,p,t) = \frac{1}{(2\pi)^2} \int d^2 u 
      \langle x+u/2 | \rho(t) | x-u/2 \rangle e^{-i p u} .
\ee
Note that $W(x,p,t)$ is still an operator in the Hilbert space of the
internal atomic degrees of freedom. However, for our specific choice of
an atomic $F=1/2$ to $F'=3/2$ transition and the laser polarizations
always lying within the same plane, no coherences between the ground-state
sublevels build up. Hence the Wigner operator $W(x,p,t)$ remains diagonal,
\bea
W(x,p,t) &=& W_+(x,p,t) |m=1/2\rangle \langle m=1/2| \nonumber \\
         &+& W_-(x,p,t) |m=-1/2\rangle \langle m=-1/2| .
\eea
For these diagonal terms we obtain the Fokker-Planck equations
\bea
   \dot{W}_\pm + \frac{p_i}{m} \partial_i W_\pm  & = & 
      -\gamma_{\pm\mp} W_\pm +\gamma_{\mp\pm} W_\mp \nonumber \\
      &+&F_{\pm\pm}^i \partial_{p_i} W_\pm + F_{\mp\pm}^i 
          \partial_{p_i} W_\mp
         \nonumber \\
      &+& 
      D_{\pm\pm}^{ij} \partial_{p_i}\partial_{p_j} W_\pm 
      +D_{\mp\pm}^{ij} \partial_{p_i}\partial_{p_j} W_\mp ,
\label{eq:fokker}
\eea
where $i,j=x,y$ and the sum over $i$, $j$ must be performed. The full
expressions for the jump rates $\gamma_{\pm\mp}$ between the ground-state
sublevels, the force coefficients $F_{\pm\pm}^i$ and the diffusion
coefficients $D_{\pm\pm}^{ij}$ are given in the appendix.

Finally, we use semiclassical Monte-Carlo simulations \cite{Castin}
to obtain numerical solutions of Eqs.~(\ref{eq:fokker}), where one follows
the trajectories of many
particles with internal states $|+\rangle$ and $|-\rangle$. The jump rates
between these two states are given by $\gamma_{+-}$ and $\gamma_{-+}$,
respectively. Between two jumps the particles evolve according to the force
$F_{++}^i$ (resp.\ $F_{--}^i$) acting on them and in addition receive random
kicks which are chosen in such a way as to simulate the effect of the
diffusion terms $D_{++}^{ij}$ (resp.\ $D_{--}^{ij}$). 

Averaging over a set of particles and over time yields all the required
expectation values such as temperatures, mean local velocities, position
distributions or spatial diffusion coefficients. We will discuss the most
important results obtained in that way in Sec.~\ref{sec:numerics}.

\subsection{Estimate of final temperature}

Before turning to the numerical results obtained from solving the
equations of motion presented in the previous subsection by Monte-Carlo
simulations, we will analytically derive a rough estimate for the
steady-state temperature $T$ obtained as the ratio of the mean momentum
diffusion coefficient $\overline{D}$
over the mean friction coefficient $\overline{\alpha}$,
\be
   k_B T=\frac{\overline{D}}{\overline{\alpha}},
\label{eq:temp3}
\ee
where the bar over $D$ and $\alpha$ denotes averaging over the internal
atomic state {\em and\/} over position. For simplicity we will restrict
the following calculations only to the $x$ direction.

In order to calculate the friction coefficient $\overline{\alpha}$ we
must find the stationary solution $W(x,p)$ of Eq.~(\ref{eq:fokker}) up
to first order in the atomic velocity $v=p/m$, i.e., we expand $W(x,p)$ by
\be
   W(x,p) = W^0(x) + \frac{p}{m} W^1(x) + \dots 
\ee
and insert this into Eq.~(\ref{eq:fokker}). Since we are considering an
atom moving with constant velocity we may neglect the force and diffusion
terms on the right-hand side of the equation and thus obtain the results
\bea
   W_+^0 = 1 - W_-^0 & = & \frac{\gamma_{-+}}{\gamma_{-+} + \gamma_{+-}}, \\
   W_+^1 = -W_-^1 & = & -\frac{\partial_x W_+^0}{\gamma_{-+} + \gamma_{+-}}.
\eea
The friction coefficient, i.e.\ the term of first order in the velocity $v$
of the force, averaged over the internal atomic state is then given by
\be
   \alpha = F_{++}^x W_+^1 + F_{--}^x W_-^1 .
\label{eq:fric2}
\ee

Finally this must be averaged over the position within the speckle field.
Since $F_{\pm\pm}^x$ as well as $W_{\pm}^1$ can be expressed in terms of
the speckle electric field amplitude and its spatial derivatives (see 
appendix), we need to know the 
expectation values of products of these quantities when averaged over
position. Considering the complete randomness of the speckle field we
assume that all averages over such products vanish except for
\be
\overline{E_+E_+^*} = \overline{E_-E_-^*} = 1,
\ee
assuming that the speckle field and the counterpropagating plane wave have
the same average intensity, and
\bea
  \overline{\partial_x E_+\partial_x E_+^*} &=& 
  \overline{\partial_x E_-\partial_x E_-^*} \nonumber \\
  & = &
\frac{1}{2} k^2 \left(\frac{\lambda}{d_{sp}}\right)^2 \overline{E_+E_+^*} = 
\frac{1}{2} k^2 \left(\frac{\lambda}{d_{sp}}\right)^2,
\label{eq:derivs}
\eea
where again $d_{sp}$ is the mean speckle grain size. The right-hand side
of Eq.~(\ref{eq:derivs}) is obtained under the assumption that
the required quantity for the speckle field is the same as for a periodic
electric field with the same typical length scale, i.e.\ for an electric
field given by
$E(x)=\cos[kx(\lambda/d_{sp})]$. Applying these assumptions
to Eq.~(\ref{eq:fric2}) yields the friction coefficient
\be
   \overline{\alpha} = \hbar k^2 \frac{3\Delta}{4\Gamma}
      \left(\frac{\lambda}{d_{sp}}\right)^2.
\label{eq:fric3}
\ee

The averaged diffusion coefficient has two different components. One of
these is obtained by averaging the diffusion coefficients 
$D_{\pm\pm}^{xx}$ over the internal atomic states and over position
similarly as done above for the friction coefficient. The second
contribution to the total diffusion coefficient arises from the 
change of the dipole force, if the atom changes its internal state. This
dipole diffusion term can be derived easily following the lines of
Ref.~\cite{DalibardCohen}. We will omit all the calculational steps here
and only give the final result for the total momentum diffusion coefficient
\be
  \frac{\overline{D}}{\hbar^2 k^2 \gamma'} = 
    \frac{1}{2}\left(\frac{\Delta}{\Gamma}\right)^2 
    \left(\frac{\lambda}{d_{sp}}\right)^2
    + \frac{11}{36}\left(\frac{\lambda}{d_{sp}}\right)^2
    + \frac{5}{36},
\label{eq:diff}
\ee
where the first term is the dipole term and the second and third terms are
due to the momentum diffusion of an atom in a definit, internal state,
i.e., coming from $D_{\pm\pm}^{xx}$.

Thus we obtain for the estimated steady-state temperature (\ref{eq:temp3})
\be
  \frac{k_B T}{\hbar\Delta'} = \frac{2}{3}
    + \frac{11}{27}\left(\frac{\Gamma}{\Delta}\right)^2
    + \frac{5}{27}\left(\frac{\Gamma}{\Delta}\right)^2
         \left(\frac{d_{sp}}{\lambda}\right)^2.
\label{eq:temp4}
\ee
We see that the expressions derived in
Eqs.~(\ref{eq:fric3})-(\ref{eq:temp4}) are essentially the same as the
intuitive ones of Sec.~\ref{sec:qualitative}.

It should be emphasized again that all quantities derived in this subsection
should be considered as crude estimates since they are performed in 1D and
rely on some very rough approximations. However, the qualitative
behavior of the exact solutions is predicted correctly and thus these
expressions provide a lot of physical insight.


\section{Numerical results}
\label{sec:numerics}

\subsection{Steady-state temperatures}

First, let us continue the discussion of the numerically found steady-state
temperature depending on various system parameters.

In Fig.~\ref{fig:detuning} we plot the steady-state temperature versus the
detuning $\Delta$ for fixed optical potential depth $\Delta^\prime$, i.e.,
for varying optical pumping rate $\gamma^\prime$. The two lower curves
correspond to the same optical potential depth but to different speckle
grain size $d_{sp}$. In agreement with Eq.~(\ref{eq:temp4}) the temperature
increases for smaller detunings $\Delta$, i.e.\ larger pumping rates 
$\gamma^\prime$. For very large detunings, corresponding to small values
of $\gamma^\prime$, the temperature assumes a constant value, which is the
same for different speckle sizes as long as the optical potential depth 
$\Delta^\prime$ is the same. On the other hand for fixed speckle size but
different values of $\Delta^\prime$ the temperatures achieved in the limit
of small values of $\gamma^\prime$ differ (cf.\ also Fig.~\ref{fig:temp}).

Fig.~\ref{fig:detuning2} shows the steady-state temperature versus the
optical potential depth $\Delta^\prime$ for fixed optical pumping rate
$\gamma^\prime$, similar to Fig.~\ref{fig:temp} but for different
parameters. Again we find a drastic increase of the temperature towards
small values of $\Delta^\prime$. In this parameter limit the total
momentum diffusion (\ref{eq:diff}) reduces to its contribution stemming
from the random recoil of the spontaneously emitted photons and hence
becomes constant, but simultaneously the Sisyphus cooling vanishes, i.e.,
the friction force (\ref{eq:fric3}) approaches zero. Hence the
diffusion starts to predominate the cooling effect and the temperature
increases.

In the opposite limit of large values of $\Delta^\prime$ the total momentum
diffusion is dominated by the dipole diffusion term of Eq.~(\ref{eq:diff}) 
and hence the temperature increases linearly with $\Delta^\prime$. 
Note on Fig.~\ref{fig:detuning2} that the steady-state temperature finally
becomes independent of the optical pumping rate $\gamma^\prime$ {\em and\/}
of the speckle size $d_{sp}$ in this limit, in agreement with 
Eq.~(\ref{eq:temp4}).

\subsection{Local properties}

In this subsection we will discuss some of the localization properties of
the cold atoms. To this end we give a contour plot of the steady-state
atomic density $\rho$ in Fig.~\ref{fig:local}(a), where the calculations
have been performed for the speckle field depicted in 
Fig.~\ref{fig:speckle}. Comparing these two figures one can see easily that
the atomic density is strongly correlated with the speckle field intensity
$I$. This statistical dependence can be demonstrated
more clearly by calculating the covariance, defined as 
\be
   \mbox{cov}(\rho,I) = 
   \langle \rho I\rangle /\sqrt{\langle \rho^2 \rangle \langle I^2 \rangle}.
\ee
We find that this quantity assumes values close to its maximum value of
one, e.g.\ for Fig.~\ref{fig:local}(a) $\mbox{cov}(\rho,I) = 0.92$. 
This means that the
intensity distribution of the speckle field is efficiently mapped onto the
atomic density distribution.

However, if one looks closer at Figs.~\ref{fig:speckle}, \ref{fig:local}(a),
one notes some discrepancies, for instance some of the relatively shallow
optical potential wells at the top of the figures are surprisingly strongly
populated. Although these features cannot be explained quantitatively, we
will describe some of the physical mechanisms in the following.

Basically these local differences are caused by the fact that the field
amplitude and the phase of the speckle field are essentially uncorrelated.
Whereas in usual optical lattices the light amplitude and phase are always
correlated, which e.g.\ in a 3D lin$\perp$lin setup yields that the optical 
potential minima coincide with places of pure circular polarization
\cite{Lattices1}, this is
not the case in our setup. Consequently the local polarization of the total
light field at places of maximum local speckle field intensity can be linear
as well
as circular, but in general will be an arbitrary elliptical polarization.
Hence only speckle grains (regions of high speckle field intensity) of mainly
circular polarization contribute to the Sisyphus-type cooling, but speckle 
grains of dominating linear polarization do not.

It should be emphasized that in speckle grains of mainly circular
polarization {\em local cooling\/} can be found, that is, even within a
single well an atom can be cooled by optical pumping
processes between the widely differing optical potentials of the two
ground-state sublevels. This is in contrast to the plane-waves 1D 
lin$\perp$lin configuration, where an atom must travel across several
potential wells in order to get cooled ({\em nonlocal cooling}).
Thus, the atom is more likely to be cooled and trapped within circularly
polarized speckle grains.

Another reason for the different steady-state atomic densities in
potential wells of the same depths is formed by the radiation pressure
force. As already mentioned above, the light field amplitude and phase of
the speckle field are mutually independent and thus also the phase gradient
can differ strongly at the bottom of the potential wells. Hence, within
some potential wells the atom experiences a relatively strong radiation
pressure force and is thus pushed away. As an example, 
Fig.~\ref{fig:local}(b) shows again a contour plot of the
speckle field intensity (same as Fig.~\ref{fig:speckle}), together with the
vector field of the local radiation pressure force. This shows, e.g., a
significant radiation pressure in the $x$ direction in the high intensity
speckle grains at the center of the figure. Correspondingly, we have
numerically found nonvanishing mean atomic velocities in these regions, which
means that there exists a stationary flow of atoms following roughly the
direction of the local radiation pressure force. This might suggest that a
description of speckled lattices in terms of fluid mechanics formalism may
be interesting to investigate, but, however, this is out of the scope of
this paper.

Finally we will briefly discuss the atomic density distribution $P(\rho)$,
i.e., the probability of finding a certain density.
A histogram plot for the speckle field intensity $P(I)$ and the
atomic density is given in Fig.~\ref{fig:dist}.

The numerically obtained speckle field intensity distribution, 
Fig.~\ref{fig:dist}(a), follows an exponential law, as theoretically 
discussed in \cite{Goodman}, with equal mean and standard deviation, i.e.,
\be
P(I/\langle I\rangle) = \exp(-I/\langle I\rangle).
\ee
As can be seen from Fig.~\ref{fig:dist}(b) the atomic density does not follow
the same law, since there is always a background of unbound atoms and hence
the probability of finding a density below a certain threshold vanishes.
But above this threshold the density distribution function closely
resembles the speckle intensity distribution, which again indicates a
strong correlation between the light field intensity and the atomic density.

\subsection{Spatial diffusion}

Another important quantity for characterizing the properties of the cooled
and trapped atoms is the spatial diffusion coefficient, which roughly
quantifies the transfer of atoms between several potential wells and hence
of the spreading of an initially small atomic cloud.

Experimentally, as well as in our Monte-Carlo simulations, the spatial
diffusion coefficient $D_s$ is obtained from the spreading of an initial
atomic cloud whose variance follows a linear law in the long-time limit,
\be
   [\Delta r(t)]^2 = [\Delta r(0)]^2 + 2 D_s t.
\ee

An estimate for the value of $D_s$ can be derived by using
\be
   D_s = \frac{\overline{D}}{\overline{\alpha}^2}
\ee
for the spatial diffusion coefficient for a Brownian motion which in certain
limits is a good approximation for Sisyphus cooling \cite{Hodapp}. Applying
our findings for the friction coefficient (\ref{eq:fric3}) and the momentum
diffusion (\ref{eq:diff}) we obtain
\bea
   D_s &=& \frac{8}{9} \frac{\gamma'}{k^2}
         \left( \frac{d_{sp}}{\lambda} \right)^2 \nonumber \\
       & \times &
       \left[
          1 + 
	  \frac{11}{18} \left( \frac{\Gamma}{\Delta} \right)^2 +
	  \frac{5}{18} \left( \frac{\Gamma}{\Delta} \right)^2
	     \left( \frac{d_{sp}}{\lambda} \right)^2
       \right].
\label{eq:spatdiff}
\eea

We can now compare this with the spatial diffusion coefficients obtained by
semiclassical Monte-Carlo simulations as depicted in
Fig.~\ref{fig:spd_det} as a function of the detuning $\Delta$. It follows
from Eq.~(\ref{eq:spatdiff}) that for 
$\gamma^\prime /\Delta^\prime = \Gamma /\Delta \ll 1$ the first term
will dominate and hence the spatial diffusion coefficient scales linearly
with $\gamma^\prime$, i.e.\ like $1/\Delta$ if $\Delta^\prime$ is kept
constant. This is in qualitative agreement with the curves of
Fig.~\ref{fig:spd_det} for not too large values of the detuning $\Delta$. 
However, Eq.~(\ref{eq:spatdiff}) only holds if the atom moves only a fraction
of the speckle size $d$ between two optical pumping processes \cite{Hodapp},
which explains the deviation of the numerically obtained curves from the
$1/\Delta$ law for large values of $\Delta$.

Note that for $\Delta\gg\Gamma$ and within its range of validity 
Eq.~(\ref{eq:spatdiff}) predicts an increase of the spatial diffusion
coefficient with the square of the speckle grain size.
This has been nicely confirmed by our numerical simulations. For an actual
experiment, where the speckle size is typically of the order of tens of
wavelengths, this means that the spatial diffusion of the atoms in the
speckle field can easily be orders of magnitude larger than for a standard
optical lattice, even if the steady-state temperatures (\ref{eq:temp4})
are comparable.

\section{Influence of the longitudinal direction}
\label{sec:longitudinal}

Compared to the transverse directions, which have been discussed so far, the
physical properties of the system along the laser propagation direction are
completely different. First, it should be emphasized again that the size
of the speckle grains in this direction is generally much larger, i.e., 
by a factor on the order of $d_{sp}/\lambda$.
But superimposed on this relatively large length scale there is a
standard 1D lin$\perp$lin configuration with a typical length scale of
half an optical wavelength, such that the system must be characterized in 
this direction by two widely different length scales. Second, this
superimposed lin$\perp$lin configuration leads to efficient cooling in the
longitudinal direction at regions in space where the
speckle field intensity approximately equals the intensity of the
counterpropagating plane wave. However, at locations where the two beam
intensities differ it also leads to a non-balanced radiation
pressure force driving the atoms away from these locations. Note that the
situation is similar to what can be found in quasi-periodic lattices 
\cite{Quasicrystals}.

Hence, the full 3D situation is much more intricate than the transverse 2D
scheme. Also numerically a complete 3D treatment would be very demanding.
Instead we performed some 2D simulations of the cooling dynamics
including the longitudinal and one transverse direction, which should give 
some relevant information about the actual behavior of the system in 3D.

In Fig.~\ref{fig:flong} we depict the temperatures in the longitudinal
direction and in the transverse direction obtained from such 2D Monte-Carlo
simulations. We see that the transverse temperatures are always higher than
the longitudinal ones. Even if there exists a strong radiation pressure in
the longitudinal direction in regions of differing speckle and plane wave
intensity, this effect is over-compensated by the much more efficient
Sisyphus cooling in this direction.

Secondly, let us remark that the transverse temperatures are larger for this
type of Monte-Carlo simulations than those obtained from simulations 
including two transverse directions. Hence the inclusion of the longitudinal
direction tends to reduce the efficiency of the transverse cooling. This
can be understood by the following arguments. Since the transverse cooling
relies on a Sisyphus effect, it is more efficient for deeper potential wells
because in this case a single quantum jump between the two adiabatic
potentials reduces the atomic energy by a larger amount. 
Thus the cooling mechanism is most effective in regions of maximum speckle
field intensity. But in the longitudinal direction, these regions correspond
exactly to the regions of maximum radiation
pressure, and hence the atom is efficiently pushed away. For the transverse
cooling this means that the atom avoids the deepest potential wells, which 
consequently reduces the cooling efficiency and therefore increases the
steady-state temperature.

For the spatial diffusion coefficient we found a much larger difference
between the longitudinal and the transverse direction than for the
temperature. This agrees well with our previous results that the leading
term in the expression of the temperature (\ref{eq:temp4}) is independent
of the speckle size whereas the spatial diffusion (\ref{eq:spatdiff})
scales with the square of the speckle size.
Hence the difference in the typical length scale between the longitudinal
direction ($\lambda$) and the transverse direction ($d_{sp}$) yields widely
different spatial diffusion. Again we note that the values for the transverse 
spatial diffusion obtained by the Monte-Carlo simulations with the
longitudinal and one transverse direction lies above the value obtained
from simulations restricted to the transverse plane, according to the
reduced cooling efficiency as discussed above.

\section{Conclusions}

In this work we have demonstrated that 3D laser cooling of neutral atoms
can be achieved by a 1D setup, if one of the two counterpropagating laser
beams of a plane-waves 1D optical molasses \cite{Pgrad12b} is
replaced by a speckle laser field. The cooling mechanism in this case
relies on a Sisyphus effect similar to the one known in usual (periodic)
lattices \cite{Polgrad} in one, two, and three dimensions. We have
calculated steady-state temperatures similar to those obtained for
periodic lattices for large red detunings of the lasers from the atomic
resonance. In this far-detuned limit the steady-state
temperatures also become independent of the mean speckle grain size.

In contrast to usual lattices we have found various local effects,
such as local cooling and nonvanishing radiation pressure forces which
are essentially decoupled from the local optical potential depth.
Furthermore, according to the speckle field the spatial diffusion of the
atoms is increased by a factor of
approximately the square of the typical speckle grain size. 

Thus for an experimental realization of this scheme two points should be
emphasized. First, the experimental setup is relatively simple since a
1D configuration is already sufficient to yield 3D sub-Doppler cooling.
On the other hand, the obtained speckle grain size is typically much larger
than an optical wavelength which gives rise to longer cooling times and a
huge increase of the spatial diffusion. Thus, the lifetime of the trapped
atomic cloud will be much shorter than in usual lattices and the
achievable atomic densities much lower. However, first experimental
results \cite{Letter} have already confirmed most of the theoretical
predictions presented in this work and reasonable agreement with the
numerically obtained values for, e.g., the steady-state temperatures has
been found.

Although such investigations and possible extensions of the scheme, e.g.,
to the study of speckled dark lattices, already present interesting results
for laser cooling, we think that a main interest lies in the possibility
of creating a disordered sample of cold atoms with well controlled
statistical characteristics. Because of the growing interest in the study
of disordered materials,
the presented system may thus be useful in this domain.

\acknowledgments

This work was supported by the European TMR Network on Quantum Structures
(contract number FMRX-CT96-0077). 

\begin{appendix}

\section{The Fokker-Planck equation}

In the following we give the exact expressions for the coefficients which
appear in the Fokker-Planck equation (\ref{eq:fokker}) as obtained from
the Wigner transform of the master equation (\ref{eq:master}) (the
derivation is straightforward but lengthy and will thus be omitted). The
jump rates read
\be
   \gamma_{\pm\mp} = \frac{2}{9}\gamma^\prime |E_\mp|^2,
\ee
the force coefficients are
\bea
   F_{\pm\pm}^i & = & \Delta^\prime \left\{
       (\partial_i E_\pm) E_\pm^\dagger
       + \frac{1}{3} (\partial_i E_\mp) E_\mp^\dagger + c.c.
       \right\} \nonumber \\
     & &   + \gamma^\prime \frac{i}{2} \left\{
       (\partial_i E_\pm) E_\pm^\dagger
       + \frac{1}{9} (\partial_i E_\mp) E_\mp^\dagger - c.c.
       \right\} , \\
   F_{\mp\pm}^i & = & \gamma^\prime \frac{i}{9} \left\{
       (\partial_i E_\pm) E_\pm^\dagger - c.c.
       \right\},
\eea
where $i=x,y$, and the diffusion coefficients are
\bea
   D_{\pm\pm}^{xx} & = & \gamma^\prime \frac{1}{8} \Big\{
      4 (\partial_x E_\pm)(\partial_x E_\pm^\dagger)
      +\frac{2}{9}(\partial_x \partial_x E_\mp) E_\mp^\dagger
         \nonumber \\
      & &+\frac{2}{9} E_\mp (\partial_x \partial_x E_\mp^\dagger)
      +\frac{8}{9} (\partial_x E_\mp)(\partial_x E_\mp^\dagger)
      \Big\} \nonumber \\
    & &  + \gamma^\prime \frac{k^2}{8}|E_\pm + \frac{1}{3} E_\mp|^2
      , \\
   D_{\pm\pm}^{yy} & = & \gamma^\prime \frac{1}{8} \Big\{
      4 (\partial_y E_\pm)(\partial_y E_\pm^\dagger)
      +\frac{2}{9}(\partial_y \partial_y E_\mp) E_\mp^\dagger
         \nonumber \\
      & & +\frac{2}{9} E_\mp (\partial_y \partial_y E_\mp^\dagger)
      +\frac{8}{9} (\partial_y E_\mp)(\partial_y E_\mp^\dagger)
      \Big\} \nonumber \\ 
    & &  + \gamma^\prime \frac{k^2}{8}|E_\pm - \frac{1}{3} E_\mp|^2
      , \\
   D_{\pm\pm}^{xy} & = & D_{\pm\pm}^{yx} = \gamma^\prime \frac{1}{8}
      \Big\{
      2 (\partial_x E_\pm)(\partial_y E_\pm^\dagger)
      +\frac{2}{9}(\partial_x \partial_y E_\mp) E_\mp^\dagger
         \nonumber \\ 
      & & +\frac{4}{9} (\partial_x E_\mp)(\partial_y E_\mp^\dagger) + c.c.
      \Big\} 
      , \\
   D_{\mp\pm}^{ii} & = & -\gamma^\prime \frac{1}{36} \Big\{
      (\partial_i \partial_i E_\pm) E_\pm^\dagger
      +E_\pm (\partial_i \partial_i E_\pm^\dagger)
         \nonumber \\ 
      & & -2 (\partial_i E_\pm)(\partial_i E_\pm^\dagger)
      \Big\} 
      + \gamma^\prime \frac{k^2}{18} |E_\pm|^2
      , \\
   D_{\mp\pm}^{xy} & = & D_{\mp\pm}^{yx} = -\gamma^\prime \frac{1}{36}
      \Big\{
      (\partial_x \partial_y E_\pm) E_\pm^\dagger
         \nonumber \\ 
      & & +(\partial_x E_\pm)(\partial_y E_\pm^\dagger) + c.c.
      \Big\} . 
\eea
For the derivation of the diffusion coefficients we have simplified the
spontaneous emission pattern by assuming that fluorescence photons are only
emitted along the $x$, $y$, and $z$ axes \cite{Castin}.

Two other points are worth a comment here concerning the application of
these diffusion constants and forces in the semiclassical Monte-Carlo
simulations outlined in Sec.~\ref{sec:semiclass}. First, it turns out that
for all realistic choices of the system parameters the cross terms 
$F_{\mp\pm}^i$ and $D_{\mp\pm}^{ij}$ between the two ground-state
sublevels can be neglected as compared with the terms $F_{\pm\pm}^i$
and $D_{\pm\pm}^{ij}$. Second, at some positions in space the diffusion
coefficients assume negative values. This property indicates a purely
quantum feature of the system, i.e., it shows that there exist positions
in space where the atomic wavefunction collapses rather than spreads out
\cite{Berg}. However, this feature {\em cannot} be mimicked by the
semiclassical simulations, where negative diffusion makes no sense.
Hence, whenever this situation occurs in our numerical simulations we set
the diffusion coefficients equal to zero to avoid this problem.

\end{appendix}

\begin{figure}[tb]
\caption{Contour plot of the intensity of a numerically created speckle
field on a grid of $64\times 64$ points.}
\label{fig:speckle}
\end{figure}

\begin{figure}[tb]
\caption{Temperature $T$ vs optical potential depth $\hbar \Delta'$ for the
lin$\perp$lin configuration (solid line) and 
the $\sigma_+ - \sigma_-$ configuration (dashed) 
for fixed optical pumping rate $\gamma' = 6 \, \omega_R$. The average speckle
grain size is $d_{sp} = 3.8 \, \lambda$.}
\label{fig:temp}
\end{figure}

\begin{figure}[tb]
\caption{Temperature vs speckle grain size for the lin$\perp$lin
configuration (solid line) and for the $\sigma_+ - \sigma_-$ configuration
(dashed) for
$\Delta'=200 \, \omega_R$ and $\gamma'=13.33 \, \omega_R$.
}
\label{fig:specklesize}
\end{figure}

\begin{figure}[tb]
\caption{Temperature (2D) vs detuning $\Delta$ for fixed potential depth.
Solid curve: $d_{sp}=1.9 \, \lambda$, $\Delta^\prime=200 \, \omega_R$, 
dashed curve: $d_{sp}=5.7 \, \lambda$, $\Delta^\prime=200 \, \omega_R$,
dotted curve: $d_{sp}=1.9 \, \lambda$, $\Delta^\prime=1000 \, \omega_R$.}
\label{fig:detuning}
\end{figure}

\begin{figure}[tb]
\caption{Temperature (2D) vs optical potential depth for fixed optical
pumping rate. 
Solid curve: $d_{sp}=3.8 \, \lambda$, $\gamma^\prime=6 \, \omega_R$, 
dashed curve: $d_{sp}=7.6 \, \lambda$, $\gamma^\prime=6 \, \omega_R$,
dotted curve: $d_{sp}=3.8 \, \lambda$, $\gamma^\prime=20 \, \omega_R$.}
\label{fig:detuning2}
\end{figure}

\begin{figure}[tb]
\caption{
(a) Contour plot of the atomic density for $\Delta'=200 \, \omega_R$ and 
$\gamma'=13.33 \, \omega_R$.
(b) Contour plot of the speckle field intensity and vector field of the
local radiation pressure force. 
The size of the spatial region is $(10\lambda )^2$.}
\label{fig:local}
\end{figure}

\begin{figure}[tb]
\caption{
(a) Histogram plot of a numerically obtained speckle field intensity.
Theoretical analyses predict an exponential behavior (smooth curve).
(b) Histogram plot of the atomic density for $\Delta'=200 \, \omega_R$, 
$\gamma' = 13.33 \, \omega_R$, $d_{sp}= 1.9\, \lambda$. }
\label{fig:dist}
\end{figure}

\begin{figure}[tb]
\caption{Spatial diffusion vs detuning $\Delta$ for fixed potential depth. 
Solid curve: $d_{sp}=1.9 \, \lambda$, $\Delta^\prime=200 \, \omega_R$;
dotted curve: $d_{sp}=1.9 \, \lambda$, $\Delta^\prime=1000 \, \omega_R$;
dashed curve: $d_{sp}=5.7 \, \lambda$, $\Delta^\prime=200 \, \omega_R$. 
(For the corresponding steady-state temperatures see 
Fig.~\ref{fig:detuning}). }
\label{fig:spd_det}
\end{figure}

\begin{figure}[tb]
\caption{Steady-state temperatures in the transverse direction (solid line) 
and in the longitudinal direction (dashed) for fixed optical pumping rate 
$\gamma'=20 \, \omega_R$ versus optical potential depth. The speckle grain
size is $d_{sp}=1.8 \, \lambda$ transversally and $d'_{sp}=7.2 \, \lambda$
longitudinally.}
\label{fig:flong}
\end{figure}


\begin{thebibliography}{99}

\bibitem{elmag} P.\ Beckmann and A.\ Spizzichino, ``The Scattering of
Electromagnetic Waves from Rough Surfaces,'' (Pergamon, Oxford, 1963);
P.\ Beckmann, ``Scattering of Light by Rough Surfaces,'' in {\it Progress
in Optics VI}, edited by E.\ Wolf (North Holland, Amsterdam, 1967), 
pp.\ 55-69.

\bibitem{Goodman} J.\ W.\ Goodman, ``Statistical Properties of Laser
Speckle Patterns,'' in {\em Laser Speckle and Related Phenomena}, edited by 
J.\ C.\ Dainty (Springer-Verlag, Berlin 1975), pp.\ 9-75.

\bibitem{statistical} J.\ C.\ Dainty, ``The Statistics of Speckle
Patterns,'' in {\it Progress in Optics XIV}, edited by E.\ Wolf 
(North Holland, Amsterdam, 1977), pp.\ 1-46.

\bibitem{newtheory} Topical Issues on ``Wave Scattering from Rough Surfaces
and Related Phenomena'' in Waves in Random Media {\bf 7}, 283-520 (1997),
{\em ibid.\/} {\bf 8}, 1-158 (1998).


\bibitem{imagery} E.\ R.\ Harvey and G.\ V.\ April, Optics Lett.\ {\bf 15}, 
740 (1990).

\bibitem{roughness} U.\ Persson, Optical Engineering {\bf 32}, 3327 (1993).

\bibitem{biophysics} Y.\ Aizu and T.\ Asakura, Optics and Laser Technology 
{\bf 23}, 205 (1991).

\bibitem{Lattices1} G.\ Grynberg and C.\ Trich\'e, in {\it Coherent and
collective interactions of particles and radiation beams}, Proceedings of the
International School of Physics ``Enrico Fermi'', Course CXXXI, Varenna 1995,
edited by A.\ Aspect, W.\ Barletta, and R.\ Bonifacio (ETS Editrice, 
Pisa, 1996), p.\ 243.

\bibitem{Lattices2} A.\ Hemmerich, M.\ Weidem\"uller, and 
T.\ W.\ H\"ansch, in {\it Coherent and
collective interactions of particles and radiation beams}, Proceedings of the
International School of Physics ``Enrico Fermi'', Course CXXXI, Varenna 1995,
edited by A.\ Aspect, W.\ Barletta, and R.\ Bonifacio (ETS Editrice, 
Pisa, 1996),p.\ 503.

\bibitem{Lattices3} P.\ S.\ Jessen and I.\ H.\ Deutsch, Adv.\ At.\ Mol.\
Opt.\ Phys.\ {\bf 37}, 95 (1996).

\bibitem{Quasicrystals} L.\ Guidoni, C.\ Trich\'e, P.\ Verkerk, and
G.\ Grynberg, Phys.\ Rev.\ Lett.\ {\bf 79}, 3363 (1997).

\bibitem{Polgrad} J.\ Dalibard and C.\ Cohen-Tannoudji, J.\ Opt.\ Soc.\
Am.\ B {\bf 6}, 2023 (1989).

\bibitem{Pgrad12a} Y.\ Castin and J.\ Dalibard, Europhys.\ Lett.\ {\bf 14},
761 (1991).

\bibitem{Pgrad12b} P.\ Verkerk, B.\ Lounis, C.\ Salomon, C.\ Cohen-Tannoudji,
J.-Y.\ Courtois, and G.\ Grynberg, Phys.\ Rev.\ Lett.\ {\bf 68}, 
3861 (1992); P.\ S.\ Jessen, C.\ Gerz, P.\ D.\ Lett, W.\ D.\ Phillips,
S.\ L.\ Rolston, R.\ J.\ C.\ Spreeuw, and C.\ I.\ Westbrook
Phys.\ Rev.\ Lett.\ {\bf 69}, 49 (1992).

\bibitem{DalibardCohen} J.\ Dalibard and C.\ Cohen-Tannoudji,
J.\ Opt.\ Soc.\ Am.\ B {\bf 2}, 1715 (1985).

\bibitem{LesHouches} C.\ Cohen-Tannoudji, in {\it Fundamental Systems in
Quantum Optics}, Proceedings of the Les Houches Summer School, Session LIII, 
edited by J.\ Dalibard, J.-M.\ Raimond, and J.\ Zinn-Justin (North-Holland,
Amsterdam, 1992), pp.\ 1-164.

\bibitem{Hodapp} T.\ W.\ Hodapp, C.\ Gerz, C.\ Furthlehner, C.\ I.\ Westbrook,
W.\ D.\  Phillips, and J.\ Dalibard, Appl.\ Phys.\ B {\bf 60},
135 (1995).

\bibitem{Huntley} J.\ M.\ Huntley, Applied Optics {\bf 28}, 4316 (1989).

\bibitem{Castin} Y.\ Castin, K.\ Berg-S{\o}rensen, J.\ Dalibard, and 
K.\ M{\o}lmer, Phys.\ Rev.\ A {\bf 50}, 5092 (1994).

\bibitem{QuantSisy}  P.\ Marte, R.\ Dum, R.\ Ta\"{\i}eb, and P.\ Zoller,
Phys.\ Rev.\ A {\bf 47}, 1378 (1993).

\bibitem{Levy} S.\ Marksteiner, K.\ Ellinger, and P.\ Zoller, Phys.\ Rev.\ A 
{\bf 53}, 3409 (1996).

\bibitem{Letter} C.\ Mennerat-Robilliard, D.\ Boiron, L.\ Guidoni, 
J.\ M.\ Fournier, P.\ Horak, and G.\ Grynberg (unpublished).

\bibitem{Berg} K.\ Berg-S{\o}rensen, Y.\ Castin, E.\ Bonderup,
and K.\ M{\o}lmer, J. Phys. B {\bf 25}, 4195 (1992).

\end{thebibliography}
\end{document}